\documentclass[12pt]{article}
\usepackage{epsf}
\usepackage{amsmath}
\usepackage{graphics}

\renewcommand{\thefootnote}{\#\arabic{footnote}}
\newcommand{\gtrsim}{ \mathop{}_{\textstyle \sim}^{\textstyle >} }
\newcommand{\lesssim}{ \mathop{}_{\textstyle \sim}^{\textstyle <} }

\begin{document}

\setcounter{footnote}{0}
\begin{titlepage}

\begin{center}

\hfill hep-ph/0509121\\
\hfill TU-753\\
\hfill September 2005\\

\vskip .5in

{\Large \bf
Gravitino Production in the Early Universe and \\
Its Implications to Particle Cosmology\footnote
{Talk given at PASCOS05, Gyeongju, Korea, June 2005.}
}

\vskip .45in

{\large
Takeo Moroi
}

\vskip .45in

{\em
Department of Physics, Tohoku University\\
Sendai 980-8578, Japan \\
}

\end{center}

\vskip .4in

\begin{abstract}

Effects of the unstable gravitino on the big-bang nucleosynthesis
(BBN) and its implications to particle cosmology are discussed.  If
the gravitino mass is smaller than $\sim 20\ {\rm TeV}$, lifetime of
the gravitino becomes longer than $\sim 1\ {\rm sec}$ and its decay
may spoil the success of the standard BBN.  In order to avoid such a
problem, upper bound on the reheating temperature after the inflation
is obtained, which may be as low as $\sim 10^{5-6}\ {\rm GeV}$.  For a
successful baryogenesis with such low reheating temeprature, a
consistent scenario based on the large cutoff supergravity (LCSUGRA)
hypothesis of supersymmetry breaking, where the gravitino and sfermion
become as heavy as $\sim O(1-10\ {\rm TeV})$, is proposed.  In the
LCSUGRA, non-thermal leptogenesis can produce large enough baryon
asymmetry.  We also see that, in the LCSUGRA scenario, relic density
of the lightest superparticle becomes consistent with the WMAP value
of the dark matter density in the parameter region required for the
successful non-thermal leptogenesis.  In this case, the dark matter
density may be reconstructed with the future $e^+e^-$ linear collider.

\end{abstract}

\end{titlepage}

\renewcommand{\thepage}{\arabic{page}}
\setcounter{page}{1}
\renewcommand{\thefootnote}{\#\arabic{footnote}}
\renewcommand{\theequation}{\thesection.\arabic{equation}}

\section{Introduction}
\setcounter{equation}{0}
\label{sec:intro}

Supersymmetry (SUSY) has been attracted many attentions not only from
particle-physics point of view but also from cosmological point of
view.  Indeed, supersymmetry may give new insights into cosmological
problems to which standard model cannot answer, like the origin of the
cold dark matter, dynamics of the scalar field responsible to the
inflation, mechanism to generate the baryon asymmetry, and so on.
Thus, even in cosmology, it is expected that SUSY will play important
roles.

In order to constract viable and natural scenario of the evolution of
the universe in the framework of the supersymmetric models, there is
one serious problem caused by the gravitino, which is the superpartner
of the graviton.  Since the gravitino may have lifetime longer than
$\sim 1\ {\rm sec}$ if its mass is lighter than $\sim O(10\ {\rm
TeV})$, thermally produced gravitino in the early universe may decay
after the big-bang nucleosynthesis (BBN).  If this is the case, the
decay products of the gravitino induce electromagnetic and hadronic
shower and the high energy particles in the shower cause dissociation
of the light elements produced by the BBN.  Since the standard BBN
scenario more or less predicts light element abundances consistent
with the observations, primordial gravitino may spoil the success of
the BBN if its abundance is too large \cite{Weinberg:zq}.  Usually,
for unstable gravitino with relatively small mass, this problem
(called ``gravitino problem'') is avoided by putting upper bound on
the reheating temperature after inflation.  (For details, see
\cite{Kawasaki:2004yh,Kawasaki:2004qu,Kohri:2005wn,otherBBN} and
references therein.).\footnote
{In fact, the reheating temperature here should be understood as the
maximal temperater when the (last) radiation dominated epoch is
realized.  If some scalar field other than the inflaton dominates the
universe after the inflation, reheating temparture here is given by
the temperature at the time of the decay of such scalar particle.
Examples of such scenarios are given in, for example, \cite{curvaton}.}
As we will see, the bound is quite stringent and hence it is required
to constract a scenario of cosmological evolution consistent with such
low reheating temperature.

Here, we would like to discuss three subjects which are closely
related.  First, we review the current situation of the calculation of
the upper bound on the reheating temperature from the gravitino
problem.  Then, we propose a cosmological scneario based on large
cutoff supergravity scenario, which is consistent with the constrants
on the reheating temperature.  In this scenario, the baryon asymmetry
of the universe is explained by the non-thermal leptogenesis while the
LSP becomes a good candidate of the dark matter of the universe.
Finally, we consider a possible test of such a scenario; we point out
that the precise reconstruction of the dark matter density may be
possible in this scenario once the superparticles are produced at the
linear collider.

\section{Gravitino Problem}
\setcounter{equation}{0}
\label{sec:gravitino}

We first briefly review the gravitino production in the early universe
and its effects on the BBN
\cite{Kawasaki:2004yh,Kawasaki:2004qu,Kohri:2005wn,otherBBN}.  
Even though the gravitino is a very weakly interacting particle, it
can be produced in the early universe by the scattering processes of
the particles in the thermal bath.  Using the thermally averaged
gravitino production cross section given in
\cite{Bolz:2000fu}, the ``yield variable'' of the gravitino, which is
defined as $Y_{3/2}\equiv\frac{n_{3/2}}{s}$, is given by
\cite{Kawasaki:2004qu}
\begin{eqnarray}
    \label{eq:Yx-new}
    Y_{3/2} &\simeq&
    1.9 \times 10^{-12}
    \nonumber \\ &&
    \times \left( \frac{T_{\rm R}}{10^{10}\ {\rm GeV}} \right)
    \left[ 1
        + 0.045 \ln \left( \frac{T_{\rm R}}{10^{10}\ {\rm GeV}}
        \right) \right]
    \left[ 1
        - 0.028 \ln \left( \frac{T_{\rm R}}{10^{10}\ {\rm GeV}}
        \right) \right],
\label{Ygrav}
\end{eqnarray}
where $n_{3/2}$ is the number density of the gravitino while
$s=\frac{2\pi^2}{45}g_{*S}(T)T^3$ is the entropy density with
$g_{*S}(T)$ being the effective number of the massless degrees of
freedom at the temperature $T$, and the reheating temperature is
defined as
\begin{eqnarray}
  T_{\rm R} \equiv
  \left(
  \frac{10}{g_* \pi^2} M_*^2 \Gamma_{\rm inf}^2
  \right)^{1/4},
  \label{T_R}
\end{eqnarray}
with $\Gamma_{\rm inf}$ being the decay rate of the inflaton.  (Here,
$M_*\simeq 2.4\times 10^{18}\ {\rm GeV}$ is the reduced Planck scale.)
As one can see, the number density of the gravitino increases as the
reheating temperature becomes larger.

Since the gravitino decays with very long lifetime (if it is
unstable), it may decay after the BBN starts.  Indeed, if the
gravitino mass $m_{3/2}$ is smaller than $\sim 20\ {\rm TeV}$, its
lifetime becomes longer than $1\ {\rm sec}$ and its decay may affect
the abundances of the light elements which are synthesised by the
standard BBN processes.  In particular, since the prediction of the
standard BBN is in a reasonable agreement with the observations,
light-element abundances become inconsistent with the observations if
the abundance of the gravitino is too large.  Consequently, we obtain
the upper bound on the reheating temperature.

In \cite{Kawasaki:2004yh,Kawasaki:2004qu,Kohri:2005wn}, with a
detailed analysis of the non-standard processes induced by the
gravitino decay (in particular, the hadro- and photo-dissociations as
well as the $p\leftrightarrow n$ conversion), light-element abundances
are calculated as a function of the reheating temperature and the
gravitino mass.  In addition, in the analysis given in
\cite{Kohri:2005wn}, decay processes of the gravitino are studied in
detail.  Then, comparing the resultant light-element abundances with
the observations, upper bound on the reheating temperature is
obtained.  

\begin{figure}[t]
   \centerline{{\vbox{\epsfxsize=0.6\textwidth\epsfbox{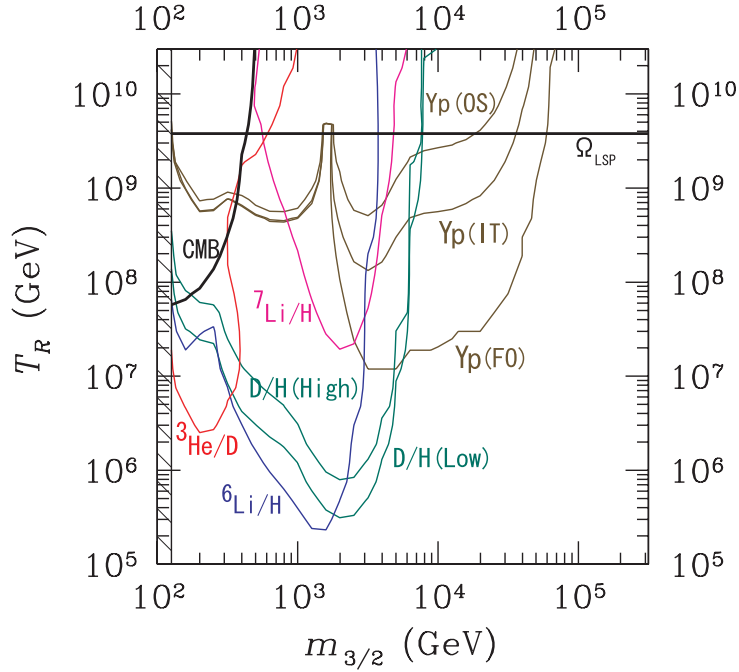}}}}
   \caption{Upper bound on the reheating temperature after inflation as 
   a function of the gravitino mass $m_{3/2}$.  The mass spectrum of the
   MSSM particles is determined by adopting the mSUGRA model with
   $m_{1/2}=300\ {\rm GeV}$,
   $m_0=2397\ {\rm GeV}$, $\mu_H=231\ {\rm GeV}$, and $\tan\beta=30$.}
   \label{fig:mtr3}
\end{figure}

One of the results is shown in Fig.\ \ref{fig:mtr3}.  Here, the
mSUGRA-type parameterisation of the minimal supersymmetric standard
model (MSSM) parameters is used and, for Fig.\ \ref{fig:mtr3}, the
following choice of the model parameters is adopted: unified gaugino
mass $m_{1/2}=300\ {\rm GeV}$, universal scalar mass $m_0=2397\ {\rm
GeV}$, SUSY invariant Higgs mass $\mu_H=231\ {\rm GeV}$, and
$\tan\beta=30$.  With this choice of parameters, the mass of the LSP
becomes $116\ {\rm GeV}$.  (The parameter region where the gravitino
mass becomes lighter than the LSP mass is shaded.)  In the figure,
each line shows the upper bound on the reheating temperature from the
considerations of different light-element abundances.  Here, the
following observational constraints are used.  For ${\rm D}/{\rm H}$
\cite{ObsD},
\begin{equation}
  {\rm D}/{\rm H} ({\rm Low})
  = (2.78^{+0.44}_{-0.38}) \times 10^{-5},
\end{equation}
and
\begin{equation}
  {\rm D}/{\rm H} ({\rm High})
    = (3.98^{+0.59}_{-0.67}) \times 10^{-5};
\end{equation}
for ${\rm ^3He}/{\rm D}$ \cite{Geiss93}
\begin{equation}
  {\rm ^3He}/{\rm D} < 0.59 \pm 0.54 ~~ (2\sigma);
    \label{He3/D}
\end{equation}
for ${\rm ^4He}$ mass fraction $Y$, one by Fields and Olive
\cite{Fields:1998gv}
\begin{equation}
    Y ({\rm FO}) = 0.238 \pm (0.002)_{\rm stat} \pm
    (0.005)_{\rm syst}, 
    \label{Y_FO}
\end{equation}
one by  Izotov and Thuan \cite{Izotov:2003xn}
\begin{equation}
    Y ({\rm IT}) = 0.242 \pm (0.002)_{\rm stat} (\pm
    (0.005)_{\rm syst}),
    \label{Y_IT}
\end{equation}
and one by Olive and Skillman \cite{Olive:2004kq}
\begin{equation}
    Y ({\rm OS}) = 0.249 \pm 0.009;
    \label{Y_OS}
\end{equation}
for $^{7}{\rm Li}$ \cite{Bonifacio:2002yx}
\begin{eqnarray}
  \label{eq:li7}
  \log_{10}
  \left({^7{\rm Li}}/{\rm H}\right)
  =-9.66 \pm (0.056)_{\rm stat} \pm (0.300)_{\rm add};
\end{eqnarray}
and for ${\rm ^6Li}$  \cite{li6_obs}
\begin{eqnarray}
    {\rm ^6Li} / {\rm H} <
    ( 1.10^{+ 5.14}_{-0.94} ) \times 10^{-11} ~~ (2 \sigma).
\end{eqnarray}

\begin{figure}[t]
   \centerline{{\vbox{\epsfxsize=0.6\textwidth\epsfbox{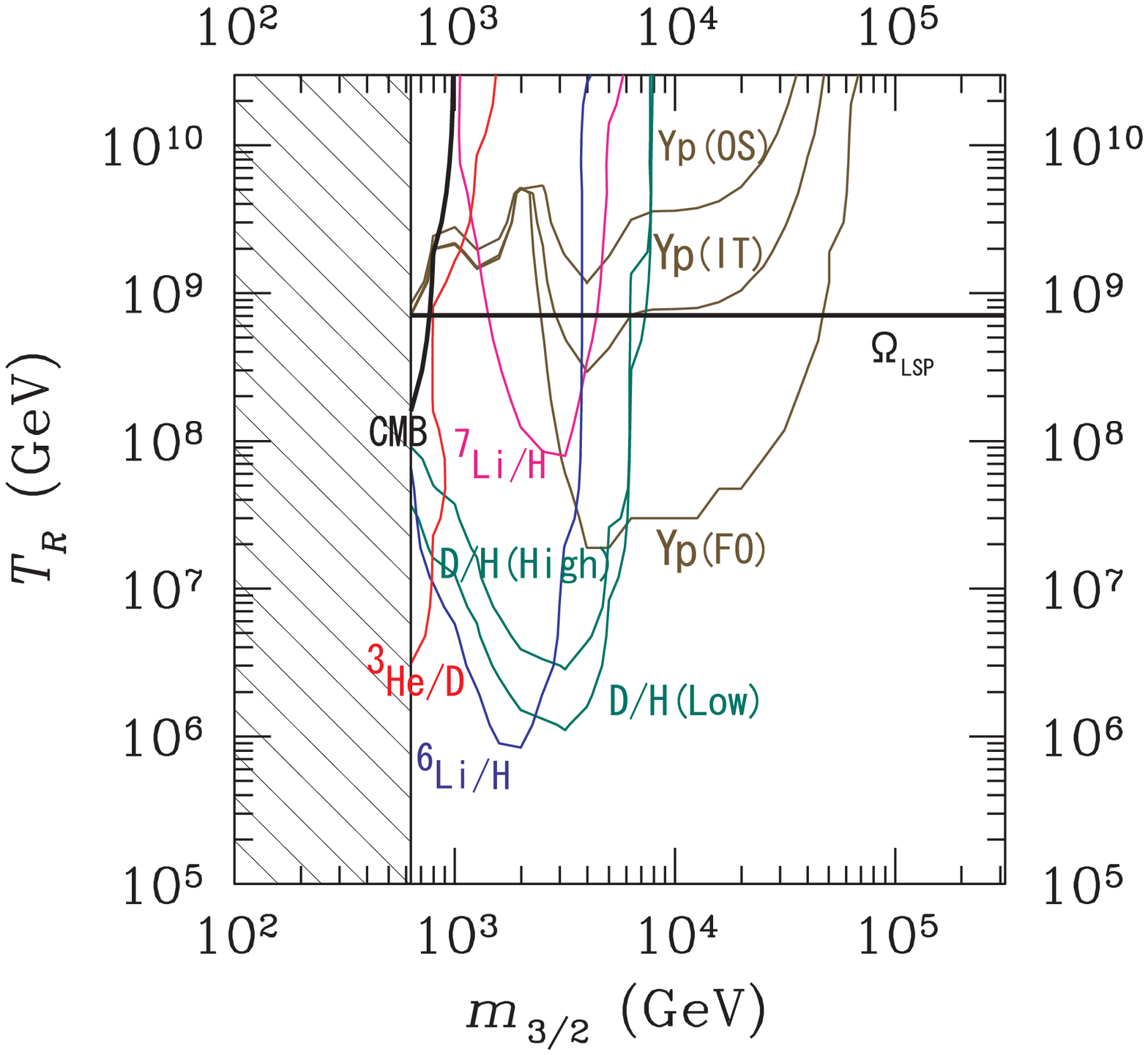}}}}
   \caption{Same as Fig.\ \ref{fig:mtr3} except
   $m_{1/2}=1200\ {\rm GeV}$,
   $m_0=800\ {\rm GeV}$, $\mu_H=1215\ {\rm GeV}$, and $\tan\beta=45$.}
   \label{fig:mtr4}
\end{figure}

The upper bound on the reheating temperature is from various effects.
When the gravitino mass is larger than a few TeV, most of the
primordial gravitinos decay at very early stage of the BBN.  In this
case, in addition, photo- and hadro-dissociations are ineffective.
Then, overproduction of ${\rm ^4He}$ due to the $p\leftrightarrow n$
conversion becomes the most important.  We note here that we consider
three different observational constraints on ${\rm ^4He}$, which are
given by Fields and Olive (FO) \cite{Fields:1998gv}, Izotov and Thuan
(IT) \cite{Izotov:2003xn} and Olive and Skillman (OS)
\cite{Olive:2004kq}.  As one can see, the upper bound on
$T_{\rm R}$ in this case is sensitive to the observational constraint
on the primordial abundance of ${\rm ^4He}$; for the case of
$m_{3/2}=10\ {\rm TeV}$, for example, $T_{\rm R}$ is required to be
lower than $3\times 10^7\ {\rm GeV}$ if we use the lowest value of $Y$
given by Fields and Olive, while, with the highest value given by
Olive and Skillman, the upper bound on the reheating temperature
becomes as large as $4\times 10^9\ {\rm GeV}$.

When $400\ {\rm GeV}\lesssim m_{3/2}\lesssim 5\ {\rm TeV}$, gravitinos
decay when the cosmic temperature is $1\ {\rm keV}$ $-$ $100\ {\rm
keV}$.  In this case, hadro-dissociation gives the most stringent
constraints; in particular, the overproduction of ${\rm D}$ and ${\rm
^6Li}$ become important.  Furthermore, when the gravitino mass is
relatively light ($m_{3/2}\lesssim 400\ {\rm GeV}$), the most
stringent constraint is from the ratio ${\rm ^3He}/{\rm D}$ which may
be significantly changed by the photo-dissociation processes of ${\rm
^4He}$.

To see how the upper bound depends on the mass spectrum of the MSSM
particles, in Fig.\ \ref{fig:mtr4}, result with $m_{1/2}=1200\ {\rm
GeV}$, $m_0=800\ {\rm GeV}$, $\mu_H=1215\ {\rm GeV}$, and
$\tan\beta=45$ is shown.  (In this case, the lightest neutralino mass
is given by $509\ {\rm GeV}$).  As one can see, the behaviour
qualitatively similar, although the detailed bound depends on the mass
spectrum.

\section{Non-Thermal Leptogenesis and LCSUGRA}
\setcounter{equation}{0}
\label{sec:lcsugra}

So far, we have seen that, in order to suppress the gravitino
production in the early universe, it is required that the reheating
temperature after the inflation be lower than $10^5 - 10^7\ {\rm GeV}$
for the gravitino mass $100\ {\rm GeV} - 10\ {\rm TeV}$, if the
gravitino is unstable.  This result imposes a serious constraint on
one of the well-motivated mechanism of baryogenesis, the leptogenesis
scenario \cite{Fukugita:1986hr} where the present baryon asymmetry of
the universe is generated from the decay of thermally produced
right-handed neutrinos.

In order to generate large enough baryon number asymmetry by the
thermal leptogenesis scenario, it is necessary to raise the reheating
temperature up to $10^{9-10}\ {\rm GeV}$ or higher \cite{thermalLG}.
We can see that such high reheating temperature conflicts with the
upper bound on $T_{\rm R}$ given in Figs.\ \ref{fig:mtr3} and
\ref{fig:mtr4} for large range of the gravitino mass.

The thermal leptogenesis may become viable if the gravitino mass is 
extremely large ($m_{3/2}\gtrsim 100\ {\rm TeV}$) or if the gravitino
is stable.  However, in this article, we would like to pursue another
direction, the non-thermal leptogenesis
\cite{NonThermalLgen}.  In the non-thermal leptogenesis
scenario, the right-handed neutrinos are assumed to be produced by the
decay of the inflaton.  Then, the rest of the scenario is almost the
same as the thermal leptogenesis; decay of the non-thermally produced
right-handed neutrinos produce lepton number asymmetry, which is
converted to the baryon number asymmetry by the spharelon effect.

One can easily see that, in the non-thermal leptogenesis scenario,
larger amount of the baryon number asymmetry can be generated with the
same reheating temperature compared to the thermal leptogenesis case.
In the non-thermal leptogenesis, the inflaton $\Phi$ decays into the
right-handed neutrinos $\Phi\rightarrow\nu_R\nu_R$.  Thus, at the time
of the reheating when the energy density of the inflaton $\rho_{\rm
inf}$ gives the rough estimate of $T_{\rm R}^4$, baryon asymmetry is
estimated by
\begin{eqnarray}
n_B (T_{\rm R}) \sim \epsilon 
\frac{\rho_{\rm inf}}{m_\Phi}.
\end{eqnarray}
Here, $\epsilon$ is the baryon asymmetry from the single decay of
$\nu_R$, and is estimated as \cite{Hamaguchi:2001gw}
\begin{eqnarray}
  \epsilon &\equiv& \frac{\Gamma (N_1\rightarrow H_u +\ell) -
    \Gamma (N_1 \rightarrow {H}_u^* + {\ell}^*)} {\Gamma _{N_1}}
  \nonumber \\ &\simeq&
  -\frac {3}{8\pi}\frac{M_1}{\langle H_u\rangle ^2}
  m_{\nu_3}\delta _{\rm eff},
  \label{epsilon}
\end{eqnarray}
where $m_{\nu_3}$ is the heaviest (active) neutrino mass, $M_1$ the
mass of the lightest right-handed neutrino, and $\delta_{\rm eff}$ the
effective CP-violating phase which is assumed to be $\sim 1$.  Using
the relation $\rho_{\rm inf}\sim T_{\rm R}^4$, we obtain the following
relation for the baryon-to-entropy ratio
\begin{eqnarray}
\frac{n_B}{s} &=&
  3.5 \times 10^{-11} 
  \nonumber \\ && \times
  \kappa
  \left(\frac{T_R}{10^6{\rm GeV}}\right)
  \left(\frac{2M_1}{m_\Phi}\right)
  \left(\frac{m_{\nu_3}}{0.05{\rm eV}}\right)
  \delta_{\rm eff},
\end{eqnarray}
where $\kappa$-parameter is a constant which is expected to be of
$O(1)$.  We evaluated this constant by numerically solving the
Boltzmann equations and found that $\kappa=2.44$.  We can see that, in
order to generate observed baryon asymmetry, reheating temperature of
$O(10^6\ {\rm GeV})$ is enough in the case of non-thermal
leptogenesis.

Of course, even if the reheating temperature is as low as $O(10^6\
{\rm GeV})$, gravitino may be still overproduced in some case, as seen
in Figs.\ \ref{fig:mtr3} and \ref{fig:mtr4}; if the hadronic branching
ratio is close to $1$, the gravitino mass should be larger than a few
TeV in this case.  Although small hierarchy is possible between the
gravitino mass and scalar masses in the observable sector even in the
gravity-mediated SUSY breaking scenario, the gravitino mass larger
than $\sim {\rm TeV}$ seems quite high from the point of view of the
naturalness of the electroweak symmetry breaking since, in the gravity
mediated SUSY breaking scenario, gravitino mass provides rough
estimate of the soft SUSY breaking scalar masses.

Recently, however, one interesting scenario has been proposed where
relatively large gravitino mass is realized.  The scenario is based on
the hypothesis such that all higher dimensional operators such as
quartic terms in the K\"ahler potential are suppressed by a cut-off
scale much higher than the Planck scale $M_*$ \cite{Ibe:2004mp}.  In
this scenario, the sfermions and gravitino become order-of-magnitude
heavier than the gauginos and, consequently, masses of the sfermions
and gravitino are required to be significantly larger than the
electroweak scale.  Even so, naturalness of the electroweak symmetry
breaking can be maintained by the focus-point mechanism \cite{focus}
due to the fact that the universality of the scalar masses at the GUT
scale is guaranteed in this scenario.  In \cite{Ibe:2004mp}, it was
shown that this scenario, called ``large cutoff supergravity (LCSUGRA)
scenario,'' is well consistent with low-energy phenomenology.  In
particular, heaviness and universality of the sfermion masses are good
for suppressing dangerous supersymmetric effects on the flavor
violating processes, proton decay, and so on.  We consider that the
presence of the large cutoff is a reflection of a more fundamental
physics beyond the GUT scale.

In the LCSUGRA scenario, it is notable that we can construct natural
and consistent scenario of cosmology \cite{Ibe:2005jf}.  In
particular, LCSUGRA predicts relatively large gravitino mass $\sim$ a
few TeV, so the serious gravitino problem may be evaded with the
reheating temperature of $O(10^{6}\ {\rm GeV})$.  As we mentioned,
even with such a low reheating temperature, baryon asymmetry can be
produced by the non-thermal leptogenesis.

In addition, it is also notable that the lightest neutralino
$\chi_1^0$ becomes the LSP in this scenario, and it can be a good
candidate of the dark matter.  Although all the sfermions acquire
multi-TeV masses in this scenario, the pair annihilation of the
lightest neutralino (which is dominantly the Bino) can be enhanced via
the sizable Higgsino component \cite{Feng:2000gh}.

\begin{figure}[t]
  \centerline{{\vbox{\epsfxsize=0.5\textwidth\epsfbox{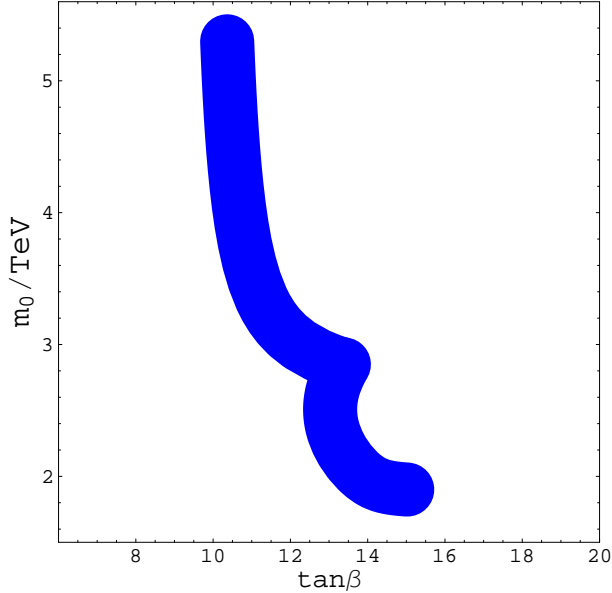}}}}
  \caption{Cosmologically allowed regions of the relic density on the
  $\tan\beta$ vs.\ $m_0$ plane for $m_{\rm top}=174$GeV.}
  \label{fig:lcsugra}
\end{figure}

The low-energy effective theory from the LCSUGRA is the same as
mSUGRA-type models, and the low-energy MSSM parameters are
parameterised by the unified gaugino mass $m_{1/2}$, the universal
scalar mass $m_0$, and $\tan\beta$.  So, we can calculate the relic
density of the LSP as a function of these parameters.  With a detailed
numerical calculations, we calculate the relic density and obtained
the parameter region where the relic density becomes consistent with
the WMAP value \cite{WMAP}
\begin{eqnarray}
\Omega_{\rm DM}^{\rm (WMAP)}h^2=0.1126^{+0.0161}_{-0.0181}.
\end{eqnarray}
The result is shown in Fig.~\ref{fig:lcsugra}.  on $\tan\beta$
vs. $m_0$ plane.\footnote
{Here, we reassure that the heavier CP-even, CP-odd Higgs bosons and
all sfermions are much heavier than the neutralino.}
In our numerical calculations, we have used the ISAJET~7.69
code~\cite{Paige:2003mg} which takes into account the one-loop
corrections to the effective Higgs potential and the two-loop RG
evolutions of parameters.\footnote
{It should be noted that, the lines in the figures show rough fitting of
the results, since the code becomes somewhat unstable for $m_0 \gg
m_{1/2}$.}
In the figure, there is an upper bound on $\tan\beta$, which is from
the lower bounds on the $\mu$ and $m_2$. The upper bound corresponds
to parameters $(\tan\beta,m_0,\mu,m_2)
\simeq (25, 4~{\rm TeV}, 140~{\rm GeV}, 160~{\rm GeV})$.
(The lower bound on the $\tan\beta$ comes from the upper bound on
$m_2\lesssim 1$~TeV where we confine our attention.)  From the figure,
we find that the relic density is consistent for the WMAP result for
$m_0\gtrsim 2$~TeV.

\section{Reconstructing $\Omega_{\rm LSP}$}
\setcounter{equation}{0}
\label{sec:recomega}

Finally, we discuss a possible test of this scenario at the future
$e^+e^-$ collider (which is recently called as the International
Linear Collider, or ILC) \cite{JLC1,NLC,Aguilar-Saavedra:2001rg}.  If
the cold dark matter consists of the LSP from the focus-point
supersymmetry, the dark matter density may be reconstructed once the
charginos and neutralinos become kinematically accessible the ILC
\cite{Moroi:2005nc}.

The important point is that, in the focus point scenario, all the
sfermions (as well as the heavier Higgses) acquire masses of $O(1\
{\rm TeV})$, so the pair annihilation of the lightest neutralino is
dominated by the processes $\chi^0_1\chi^0_1\rightarrow t\bar{t}$,
$\chi^0_1\chi^0_1\rightarrow W^+W^-$ and $\chi^0_1\chi^0_1\rightarrow
ZZ$.  These processes are via the Higgsino component in the lightest
neutralino, as mentioned before.  Thus, in this case, the dark-matter
density is primarily determined by four parameters $m_{\rm G1}$,
$m_{\rm G2}$, $\mu_H$, and $\tan\beta$.  (Here, $m_{\rm G1}$ and
$m_{\rm G2}$ are gaugino masses for the $U(1)_Y$ and $SU(2)_L$ gauge
groups, respectively.)  Since some of the superparticles (in
particular, the charginos and the neutralinos) are relatively light,
we may be able to obtain information about these parameters from the
study of these superparticles.

If the relic density of the LSP is close to the WMAP value, the LSP is
the Bino-like neutralino.  In addition, in order for the sizable
contamination of the Higgsino component into the lightest neutralino,
the $\mu_H$ parameter is required to be relatively small.
Consequently, the lightest chargino as well as the second and third
lightest neutralinos become Higgsino like.  Study of their properties
will give important information for the calculation of the thermal
relic density of the LSP.

Once the Higgsino-like charginos and neutralinos are produced at the
ILC, their properties as well as the mass of the LSP can be precisely
determined.  For example, from the threshold can at $\sqrt{s}\sim
2m_{\chi^\pm_1}$, we can determine the chargino mass from the process
$e^+e^-\rightarrow\chi^+_1\chi^-_1$.  In the linear collider,
neutralinos can be also produced.  Since the neutralinos are Majorana
particle, pair productions of the identical neutralinos are suppressed
at the threshold region.  The process
$e^+e^-\rightarrow\chi^0_2\chi^0_3$ can have, however, sizable cross
section.  From the threshold scan of this process, we can determine
the combination $m_{\chi^0_2}+m_{\chi^0_3}\equiv
2\bar{m}_{\chi^0_{23}}$.  At the ILC, errors in the measurements of
the masses are expected to be mostly from the detector resolutions
\cite{Aguilar-Saavedra:2001rg}.  For example, it was pointed out that,
for some choice of the SUSY parameters, masses of the charginos can be
determined using $e^+e^-$ colliders with the errors of $\sim 50\ {\rm
MeV}$ by the threshold scan.  In addition, from the energy
distribution of the decay products of the chargino and neutralinos,
the mass of the LSP is also determined with the uncertainty of $\sim
50\ {\rm MeV}$.  Although these results are for the case of Wino-like
chargino and neutralino, we expect that three mass parameters (i.e.,
$m_{\chi^\pm_1}$, $m_{\chi^0_1}$, and $\bar{m}_{\chi^0_{23}}$) are
accurately measured once $\chi^\pm_1$, $\chi^0_2$, and $\chi^0_3$
become kinematically accessible at the ILC.  Since $\chi^0_1$ is
Bino-like while $\chi^\pm_1$ (as well as $\chi^0_2$ and $\chi^0_3$)
are Higgsino-like, we can constrain $m_{\rm G1}$ and $\mu_H$ from the
measurements of $m_{\chi^0_1}$ and $m_{\chi^\pm_1}$ (or from the
masses of other Higgsino-like neutralinos).

Thus, if the Higgsino-like chargino ($\chi^\pm_1$) and neutralinos
($\chi^0_2$ and $\chi^0_3$) are produced at the ILC, three constraints
will be obtained on the MSSM parameters $m_{\rm G1}$, $m_{\rm G2}$,
$\mu_H$, and $\tan\beta$.  Of course, if, for example, the heavier
chargino can become kinematically accessible at the ILC, we can impose
four constraints on the MSSM parameters so all the parameters relevant
for the calculation of the $\Omega_{\rm LSP}$ can be in principle
reconstructed.  Even without producing the Wino-like chargino and
neutralino, however, interesting bound on $\Omega_{\rm LSP}$ can be
obtained.  To see this, we can perform the following analysis.  Let us
imagine a situation where $m_{\chi^\pm_1}$, $m_{\chi^0_1}$, and
$\bar{m}_{\chi^0_{23}}$ are well measured at the ILC.  Using these
quantities, we impose three constraints on the four underlying
parameters and determine $m_{\rm G1}$, $\mu_H$, and $\tan\beta$ as
functions of $m_{\rm G2}$.  In the determination of $m_{\rm G1}$ and
$\mu_H$, in fact, there are four possible choices of their signs:
$({\rm sign}(m_{\rm G1}), {\rm sign}(\mu_H))=(+,+)$, $(+,-)$, $(-,+)$,
and $(-,-)$.\footnote
{To be more precise, these signs are the relative signs between
$m_{\rm G1}$ and $m_{\rm G2}$ or $\mu_H$ and $m_{\rm G2}$.  We assume
that the gaugino masses and $\mu_H$ are real in order to avoid
constraints from CP violations.}
Effects of the signs of $\mu_H$ and $m_{\rm G1}$ are quite different.
In order to see how the reconstructed relic density depends on $m_{\rm
G2}$ and ${\rm sign}(\mu_H)$, here we consider the case where the sign
of the reconstructed $m_{\rm G1}$ is the same as that of the
underlying one; effects of ${\rm sign}(m_{\rm G1})$ will be discussed
later.  Once we reconstruct $m_{\rm G1}$, $\mu_H$, and $\tan\beta$, we
calculate the relic density of the LSP as a function of $m_{\rm G2}$,
which we call
\begin{eqnarray*}
    \hat{\Omega}_{\rm LSP} 
    (m_{\rm G2}; m_{\chi^\pm_1}, m_{\chi^0_1}, 
    \bar{m}_{\chi^0_{23}}).
\end{eqnarray*}

\begin{figure}[t]
  \centerline{{\vbox{\epsfxsize=0.6\textwidth\epsfbox{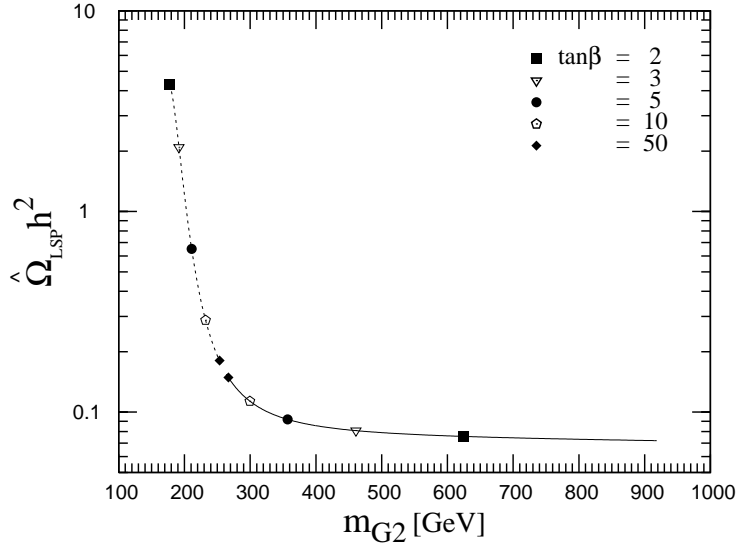}}}}
  \caption{
  $\hat{\Omega}_{\rm LSP} (m_{\rm G2}; m_{\chi^\pm_1}, m_{\chi^0_1},
  \bar{m}_{\chi^0_{23}})$ as a function of $m_{\rm G2}$, where
  $m_{\chi^\pm_1}$, $m_{\chi^0_1}$, and $\bar{m}_{\chi^0_{23}}$ are
  fixed by the underlying values for Point 1 with positive $\mu_H$
  (solid) and negative $\mu_H$ (dashed).  Marks on the figure indicate
  the points with $\tan\beta=2,3,5,10,50$.}  \label{fig:omega1}
\end{figure}

\begin{figure}[t]
  \centerline{{\vbox{\epsfxsize=0.6\textwidth\epsfbox{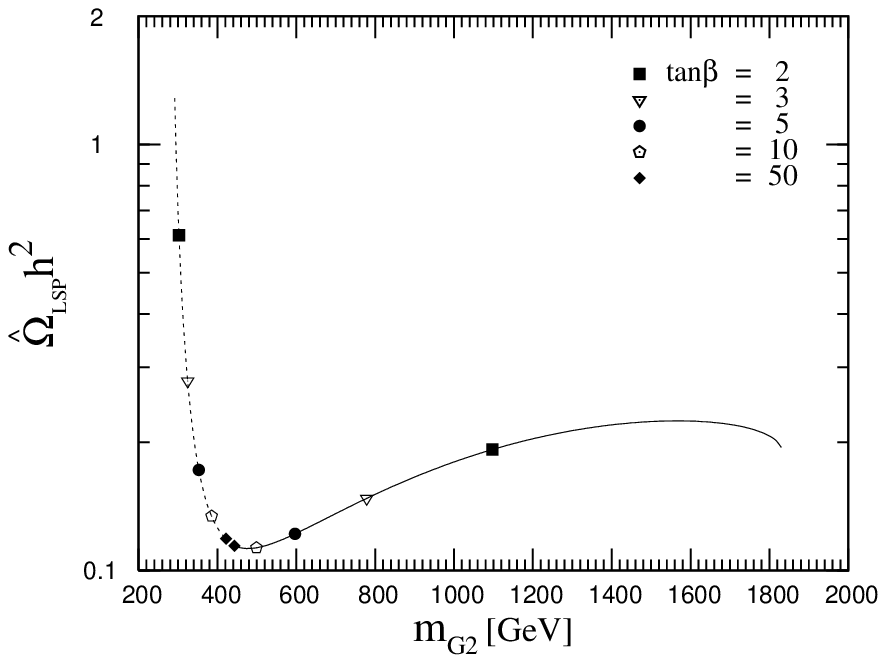}}}}
  \caption{Same as Fig.\ \ref{fig:omega1}, except for Point 2.}
  \label{fig:omega2}
\end{figure}

In Figs.\ \ref{fig:omega1} and \ref{fig:omega2}, we plot
$\hat{\Omega}_{\rm LSP} (m_{\rm G2}; m_{\chi^\pm_1}, m_{\chi^0_1},
\bar{m}_{\chi^0_{23}})$ as a function of $m_{\rm G2}$ for two 
different choices of the underlying Points: Point 1 with $m_{\rm
G1}=144\ {\rm GeV}$, $m_{\rm G2}=300\ {\rm GeV}$, $\mu_H= 200\ {\rm
GeV}$, and $\tan\beta=10$, and Point 2 with $m_{\rm G1}=240\ {\rm
GeV}$, $m_{\rm G2}=500\ {\rm GeV}$, $\mu_H= 307\ {\rm GeV}$, and
$\tan\beta=10$.  The lines have endpoints; this is due to the fact
that, when $m_{\rm G2}$ becomes too large or too small, there is no
value of $\tan\beta$ which consistently reproduces the observed mass
spectrum.  To demonstrate this, we also showed the points where
$\tan\beta$ takes several specific values.  In the figures, the
results for the cases with positive and negative $\mu_H$ are shown.

As one can see, the case with negative $\mu_H$ may give large
uncertainty in the reconstructed $\Omega_{\rm LSP}$.  If $\mu_H<0$,
however, smaller $m_{\rm G2}$ is required than in the case of
positive-$\mu_H$ in order to reproduce the observed mass spectrum, as
shown in the figures.  Then, $m_{\chi^\pm_2}$, for example, may become
smaller than the experimental bound from the negative search for the
$\chi^\pm_1\chi^\mp_2$ production process.  For Point 1 (Point 2),
$\sqrt{s}\gtrsim 480\ {\rm GeV}$ ($750\ {\rm GeV}$) is enough to
exclude the $\mu_H<0$ case.  Thus, in the following, we assume that
this is the case and neglect the uncertainty from the $\mu_H<0$ case.
Then, even without any further constraint on $m_{\rm G2}$, relic
density of the LSP can be determined within a factor of $\sim 2$ or
smaller.  Of course, if the Wino-like chargino and neutralino become
kinematically accessible at the ILC, we can determine $m_{\rm G2}$ and
hence the relic density of the LSP can be determined with a great
accuracy.\footnote
{In such a case, radiative corrections to the masses and
cross sections should be also calculated to reduce theoretical
uncertainties  \cite{Allanach:2004xn}.}

So far, we have not considered effects of the sign of $m_{\rm G1}$.
Unfortunately, $\Omega_{\rm LSP}$ depends on the sign of $m_{\rm G1}$
although the determination of ${\rm sign}(m_{\rm G1})$ seems
challenging.  For the case with negative $m_{\rm G1}$, $\Omega_{\rm
LSP}h^2$ varies from 0.9 to 7.4 (Point 1) and from 0.2 to 6.7 (Point
2).  If ${\rm sign}(m_{\rm G1})$ is undetermined, thus, two-fold
ambiguity will remain. However, experimental determination of ${\rm
sign}(m_{\rm G1})$ may be possible \cite{Choi:2001ww}.  In addition,
if the GUT relation among the (absolute values of) gaugino masses is
experimentally confirmed, it will give another hint of the signs of
the gaugino masses.

\section{SUMMARY}

Here, it is discussed that the LCSUGRA framework can provide an
interesting cosmological scenario which explains the origin of the
baryon asymmetry of the universe as well as the identity of the dark
matter without conflicting the gravitino problem.  In this scenario,
the gravitino acquires the mass of a few TeV and hence the reheating
temperature after the inflation can be as high as $O(10^6\ {\rm
GeV})$.  With such a reheating temperature, the baryon asymmetry of
the universe can be generated by the non-thermal leptogenesis.  In
addition, the lightest neutralino becomes the LSP in this scenario and
its relic density can be consistent with the dark matter density
determined by the WMAP.  Importantly, the relic density of the
LSP can be reconstructed once the superparticles are produced at the
ILC.


\section*{Acknowledgments}

The author would like to thank M. Ibe, M. Kawasaki, K. Kohri,
Y. Shimizu, T. Yanagida, and A. Yotsuyanagi for fruitful
collaborations, based on which this presentation is organised.  The
work of T.M. is supported in part by the 21st century COE program,
``Exploring New Science by Bridging Particle-Matter Hierarchy,'' and
also by the Grants-in Aid of the Ministry of Education, Science,
Sports, and Culture of Japan No.\ 15540247.


\begin{thebibliography}{30}

\bibitem{Weinberg:zq}
    S.~Weinberg,
    Phys.\ Rev.\ Lett.\  {\bf 48} (1982) 1303.

\bibitem{Kawasaki:2004yh}
  M.~Kawasaki, K.~Kohri and T.~Moroi,
  arXiv:astro-ph/0402490.

\bibitem{Kawasaki:2004qu}
  M.~Kawasaki, K.~Kohri and T.~Moroi,
  arXiv:astro-ph/0402490;
  Phys.\ Rev.\ D {\bf 71} (2005) 083502.

\bibitem{Kohri:2005wn}
  K.~Kohri, T.~Moroi and A.~Yotsuyanagi,
  arXiv:hep-ph/0507245.

\bibitem{otherBBN} 
  For recent studies on the effects of the unstable gravitino on the
  BBN other than \cite{Kawasaki:2004yh,Kawasaki:2004qu,Kohri:2005wn},
  see, for example,
    R.~H.~Cyburt, J.~R.~Ellis, B.~D.~Fields and K.~A.~Olive,
    Phys.\ Rev.\ D {\bf 67} (2003) 103521;
    K.~Jedamzik,
    Phys.\ Rev.\ D {\bf 70} (2004) 063524;

\bibitem{curvaton}
    K.~Enqvist and M.~S.~Sloth,
    Nucl.\ Phys.\ B {\bf 626} (2002) 395;
    D.~H.~Lyth and D.~Wands,
    Phys.\ Lett.\ B {\bf 524} (2002) 5;
    T.~Moroi and T.~Takahashi,
    Phys.\ Lett.\ B {\bf 522} (2001) 215
    [Erratum-ibid.\ B {\bf 539} (2002) 303];
    T.~Moroi and T.~Takahashi,
    Phys.\ Rev.\ D {\bf 66} (2002) 063501.

\bibitem{Bolz:2000fu}
    M.~Bolz, A.~Brandenburg and W.~Buchmuller,
    Nucl.\ Phys.\ B {\bf 606} (2001) 518.

\bibitem{ObsD}
    D.~Tytler, X.~m.~Fan and S.~Burles,
    Nature {\bf 381} (1996) 207;
    S.~Burles and D.~Tytler,
    Astrophys.\ J.\  {\bf 499} (1998) 699;
    S.~Burles and D.~Tytler,
    Astrophys.\ J.\  {\bf 507} (1998) 732;
    J.~M.~O'Meara et al., 
    Astrophys.\ J.\  {\bf 552} (2001) 718;
    M.~Pettini and D.~V.~Bowen,
    Astrophys.\ J.\  {\bf 560} (2001) 41;
    D.~Kirkman et al., 
    Astrophys.\ J.\ Suppl.\  {\bf 149} (2003) 1.

\bibitem{Geiss93}
    J. Geiss,
    ``Origin and Evolution of the Elements,''
    edited by N. Prantzos, E. Vangioni-Flam, and M. Cass\'e
    (Cambridge University Press, 1993) 89.

\bibitem{Fields:1998gv}
    B.~D.~Fields and K.~A.~Olive,
    Astrophys.\ J.\  {\bf 506} (1998) 177.

\bibitem{Izotov:2003xn}
    Y.~I.~Izotov and T.~X.~Thuan,
    Astrophys.\ J.\  {\bf 602} (2004) 200.

\bibitem{Olive:2004kq}
    K.~A.~Olive and E.~D.~Skillman,
    Astrophys.\ J.\  {\bf 617} (2004) 29.

\bibitem{Bonifacio:2002yx}
    P.~Bonifacio et al.,
    Astron.\ Astrophys.\ {\bf 390} (2002) 91.

\bibitem{li6_obs}
    V.~V.~Smith, D.~L.~Lambert, and P.~E.~Nissen,
    Astrophys.\ J. {\bf 408} (1993) 262;
    L.~M.~Hobbs and J.~A.~Thorburn,
    Astrophys.\ J. {\bf 491} (1997) 772;
    V.~V.~Smith, D.~L.~Lambert, and P.~E.~Nissen,
    Astrophys.\ J. {\bf 506} (1998) 923;
    R.~Cayrel et al.,
    Astron.\ Astrophys.\ {\bf 343} (1999) 923.

\bibitem{Fukugita:1986hr}
  M.~Fukugita and T.~Yanagida,
  Phys.\ Lett.\ B {\bf 174} (1986) 45.

\bibitem{thermalLG}
    See, for example, W.~Buchmuller and M.~Plumacher,
    Int.\ J.\ Mod.\ Phys.\ A {\bf 15} (2000) 5047.

\bibitem{NonThermalLgen}
  G.~Lazarides and Q.~Shafi,
  Phys.\ Lett.\ B {\bf 258} (1991) 305.
  K.~Kumekawa, T.~Moroi and T.~Yanagida,
  Prog.\ Theor.\ Phys.\  {\bf 92} (1994) 437;
  T.~Asaka, K.~Hamaguchi, M.~Kawasaki and T.~Yanagida,
  Phys.\ Lett.\ B {\bf 464} (1999) 12.

\bibitem{Hamaguchi:2001gw}
  K.~Hamaguchi, H.~Murayama and T.~Yanagida,
  Phys.\ Rev.\ D {\bf 65} (2002) 043512.

\bibitem{Ibe:2004mp}
  M.~Ibe, K.~I.~Izawa and T.~Yanagida,
  Phys.\ Rev.\ D {\bf 71} (2005) 035005.

\bibitem{focus}
    J.~L.~Feng and T.~Moroi,
    Phys.\ Rev.\ D {\bf 61} (2000) 095004;
    J.~L.~Feng, K.~T.~Matchev and T.~Moroi,
    Phys.\ Rev.\ Lett.\  {\bf 84} (2000) 2322;
    Phys.\ Rev.\ D {\bf 61} (2000) 075005.

\bibitem{Ibe:2005jf}
  M.~Ibe, T.~Moroi and T.~Yanagida,
  arXiv:hep-ph/0502074.

\bibitem{Feng:2000gh}
  J.~L.~Feng, K.~T.~Matchev and F.~Wilczek,
  Phys.\ Lett.\ B {\bf 482} (2000) 388
  [arXiv:hep-ph/0004043].

\bibitem{WMAP}
    D.~N.~Spergel et al.,
    Astrophy.\ J.\ Suppl.\ {\bf 148} (2003) 175.

\bibitem{Paige:2003mg}
    F.E.~Paige, S.D.~Protopescu, H.~Baer, and X.~Tata,
    arXiv:hep-ph/0312045.

\bibitem{JLC1}
    S.~Matsumoto {\it et al.} [JLC Group],
    ``JLC-1,'' 
    KEK Report 92-16 (1992).

\bibitem{NLC}
    S.~Kuhlman {\it et al.} [The NLC ZDR Design Group and The NLC
    Physics Working Group],
    ``Physics and Technology of the Next Linear Collider,''
    BNL 52-502 (1996).

\bibitem{Aguilar-Saavedra:2001rg}
    J.~A.~Aguilar-Saavedra {\it et al.}  
    [ECFA/DESY LC Physics Working Group],
    arXiv:hep-ph/0106315.

\bibitem{Moroi:2005nc}
  T.~Moroi, Y.~Shimizu and A.~Yotsuyanagi,
  arXiv:hep-ph/0505252.

\bibitem{Allanach:2004xn}
    B.~C.~Allanach, G.~Belanger, F.~Boudjema and A.~Pukhov,
    JHEP {\bf 0412} (2004) 020.

\bibitem{Choi:2001ww}
    S.~Y.~Choi, J.~Kalinowski, G.~Moortgat-Pick and P.~M.~Zerwas,
    Eur.\ Phys.\ J.\ C {\bf 22} (2001) 563
    [Addendum-ibid.\ C {\bf 23} (2002) 769].

\end{thebibliography}
\end{document}